\documentclass[12pt]{article}
 \usepackage{epsfig}
 \def\be{\begin{equation}}
 \def\ee{\end{equation}}
 \def\bea{\begin{eqnarray}}
 \def\eea{\end{eqnarray}}
 \usepackage{graphicx}

 \catcode`\@=11
 \def\lsim{\mathrel{\mathpalette\@versim<}}
 \def\gsim{\mathrel{\mathpalette\@versim>}}
 \def\@versim#1#2{\vcenter{\offinterlineskip
 \ialign{$\m@th#1\hfil##\hfil$\crcr#2\crcr\sim\crcr } }}
 \catcode`\@=12

 \catcode`@=12
\begin{document}
 \thispagestyle{empty}
 \begin{flushright}
 UCRHEP-T604\\
 Nov 2020\
 \end{flushright}
 \vspace{0.6in}
 \begin{center}
 {\LARGE \bf Gauged Baryon Number and\\ Dibaryonic Dark Matter\\}
 \vspace{1.5in}
 {\bf Ernest Ma\\}
 \vspace{0.1in}
{\sl Department of Physics and Astronomy,\\ 
University of California, Riverside, California 92521, USA\\}
\end{center}
 \vspace{1.2in}

\begin{abstract}\
The minimal standard model of quarks and leptons is extended with a set 
of vectorlike fermions to allow baryon number $B$ to become a gauged $U(1)_B$ 
symmetry.  The $B$ assignments of the new particles are determined by 
renormalizable interactions with the known quarks through a color triplet 
scalar diquark.  The spontaneous breaking of $U(1)_B$ by a scalar with $B=3$ 
results in a conserved residual global $B$ symmetry.  A singlet 
neutral scalar with $B=2$ is a possible long-lived dark-matter candidate.  
\end{abstract}

\newpage
\baselineskip 24pt

\noindent \underline{\it Introduction}~:~ 
In the minimal standard model (SM) of quarks and leptons, with no right-handed 
neutrino $N_R$ and just one Higgs scalar doublet, it is well-known that baryon 
number $B$ and lepton number $L$ are automatically conserved.  If one 
additional particle is added subject only to the constraint of gauge 
invariance, how would it affect $B$ and $L$?  The most often considered 
candidate is $N_R$, in which case the allowed terms 
$\bar{N}_R (\nu_L \phi^0 - e_L \phi^+)$ and $N_R N_R$ change $L$ conservation 
to $L_P = (-1)^L$ conservation, and neutrinos are Majorana fermions.

Other particles should also be looked at, each coupling to two SM particles, 
so that its $B$ and $L$ assignments may be fixed.  In fact, an analysis of 
all such scalar bilinears was done~\cite{mrs99} already in 1998.  
Subsequently~\cite{kms02}, all possible combinations of these scalars 
were considered, regarding their effect on $B$ and $L$, subject only to 
the constraint of gauge invariance.  An application to baryogenesis was 
also proposed~\cite{mrs99-1}.  

In this work, baryon number is promoted to a $U(1)_B$ gauge 
symmetry~\cite{fw10} using a set of vectorlike fermions~\cite{dfw13} 
with nonzero $B$ values. The new requirement here is that the $B$ 
assignments of these new fermions be fixed by explicit renormalizable 
interactions with the known quarks through a color triplet scalar diquark.  
As a consequence, even though $U(1)_B$ is spontaneously broken, the 
resulting theory conserves $B$ as a global symmetry, and a singlet $B=2$ 
scalar may become a long-lived dark-matter~\cite{bh18} candidate.  The 
idea of singlet fermions and scalars with integral $B$ and $L$ values is 
not new~\cite{m17}, and there have been many recent 
studies~\cite{mt18,m19,m20,h20,fhz20}. 

Note that in this work, lepton number is not gauged.  Furthermore, 
there is no right-handed neutrino.  Small Majorana neutrino masses 
are obtained through a very heavy Higgs triplet~\cite{ms98}, enabling 
also the generation of a baryon asymmetry through leptogenesis~\cite{fy86}.

\noindent \underline{\it Gauging of Baryon Number}~:~ 
The particle content of the SM does not admit the gauging of $B$ or $L$ 
because of anomalies.  However, if one right-handed singlet neutrino is 
added per family, the combination $B-L$ is well-known to be suitable as 
a gauge symmetry because the set of fermions is now anomaly-free with 
the addition of this $U(1)_X$ to the existing 
$SU(3)_C \times SU(2)_L \times U(1)_Y$.  If $B$ and $L$ are chosen 
differently for each family, then the condition for the existence of 
$U(1)_X$ is simply~\cite{kmpz17}
\begin{equation}
\sum^3_{i=1} 3 n_i + n'_i = 0,
\end{equation}
where $n_i,n'_i$ are the $U(1)_X$ values for each quark and lepton family.  
If $n_i=1/3$ and $n'_i=-1$, then $B-L$ is obtained.  Many other choices 
have been considered, such as $L_\mu-L_\tau$~\cite{hjlv91,mrr02}, 
$B-3L_\tau$~\cite{m98}, etc.

Suppose all known three families of quarks and leptons are assigned their 
canonical values, then the separate gauging of $B$ and $L$ was shown to 
be possible~\cite{fw10} by the addition of a fourth family with $B=-1$ 
and $L=-3$.  It was subsequently realized~\cite{dfw13} that vectorlike 
fermions with $B$ and $L$ values work just as well.  Here the gauging 
of only $B$ is considered, with the following additional particle 
content, all of which are color singlets.
\begin{table}[tbh]
\centering
\begin{tabular}{|c|c|c|c|}
\hline
new fermion & $SU(2)_L$ & $U(1)_Y$ & $U(1)_B$ \\
\hline
$(E^0,E^-)_L$ & $2$ & $-1/2$ & $y$ \\ 
$(E^0,E^-)_R$ & $2$ & $-1/2$ & $y+3$ \\
\hline
$X^-_R$ & $1$ & $-1$ & $y$ \\
$X^-_L$ & $1$ & $-1$ & $y+3$ \\
\hline
$X^0_R$ & $1$ & $0$ & $y$ \\
$X^0_L$ & $1$ & $0$ & $y+3$ \\
\hline
\end{tabular}
\caption{New fermions transforming under $U(1)_B$.}
\end{table}
The triangle gauge anomalies are all zero, without regard to the value of $y$, 
as shown below.
\begin{eqnarray}
[SU(2)]^2 U(1)_B &:& {3 \over 2} [SM] + {1 \over 2}(y-y-3) = 0, \\~
[U(1)_Y]^2 U(1)_B &:& -{3 \over 2} [SM] + (2)\left( -{1 \over 2} \right)^2 
(y-y-3) + (-1)^2(y+3-y) = 0, \\ 
U(1)_Y [U(1)_B]^2 &:& (2) \left( -{1 \over 2} \right) [y^2-(y+3)^2] + 
(-1)[(y+3)^2-y^2] = 0, \\~ 
[U(1)_B]^3 &:& 2[y^3-(y+3)^3]+2[(y+3)^3-y^3] = 0.
\end{eqnarray}

The new fermions obtain masses from a complex singlet scalar $\zeta$ with 
$B=3$ independent of $y$, breaking $U(1)_B$ spontaneously.  The resulting 
theory consists of three sectors, the quarks, the leptons, and the new 
fermions.  They communicate only through the gauge bosons, the SM Higgs 
boson and one real singlet scalar.  This means that the lightest particle 
in each fermion sector is stable, allowing the lightest neutral new fermion 
to be dark matter automatically.  The choice of $y=-3/2$ was made in 
Ref.~\cite{dfw13}, but it is not unique.  To justify the interpretation 
of baryon number for the new fermions, a specific connection to the known 
quarks should be made.  To this end, color scalars play the important role 
as a bridge to the new fermions.

\noindent \underline{\it Color Scalars}~:~ 
Combining two quarks or a quark with a lepton, color triplet and sextet 
scalars may be formed~\cite{mrs99}.  If a color triplet is considerd, 
it is well-known that SM gauge invariance alone allows it to couple both 
to one quark and one lepton, as well as two antiquarks, breaking $B$ and 
$L$ separately but preserving $B-L$.  This is then a source for proton 
decay.  However, if the right-handed singlet neutrino $N_R$ is absent, 
then the color scalar diquark
\begin{equation}
\eta \sim (3^*,1,-2/3)
\end{equation}
couples only to $d_R d_R$ and carries $B=2/3$.  Hence there are two $B=1$ 
fermionic combinations, namely
\begin{equation}
\eta d_R \sim (1, 1, -1), ~~~ \eta u_R \sim (1,1,0).
\end{equation}
The former may couple to $\overline{X^-_L}$ and the latter to 
$\overline{X^0_L}$, in which case $y+3=1$, so $y$ is determined to be $-2$. 
Another possible invariant is $X^0_R \eta u_R$ in which case $y=-1$. 
After spontaneous breaking of gauge $U(1)_B$ by three units through $\zeta$, 
a global residual $B$ symmetry remains.  This idea was first explicitly 
noted in Ref.~\cite{mpr13} and applied also to Dirac neutrinos~\cite{ms14} 
from the breaking of gauge $B-L$ by three units. The $y=-2$ choice corresponds 
to having $B=1$ for all the new fermions and they decay to three quarks 
through $\eta$.  The $y=-1$ choice corresponds to having $B=-1$ for all 
the new fermions and they decay to three antiquarks through $\eta^*$.

\noindent \underline{\it New Fermions}~:~
For $y=-2$ or $y=-1$, all new fermions acquire Dirac masses through 
$\langle \zeta \rangle = u/\sqrt{2}$.  Note that if $y=-3/2$, then 
$\zeta X_R^0 X_R^0$ and $\zeta^* X_L^0 X_L^0$ would be possible, and the 
residual global symmetry of the new fermion sector would be lost, but a 
dscrete $Z_2$ symmetry remains without any connection to the quarks 
unless exotic new particles are added.  The doublet $(E^0,E^-)$ is 
connected to the singlets $X^-$ and $X^0$ through the SM Higgs doublet 
$\Phi = (\phi^+,\phi^0)$.  The four invariant Yukawa terms are
\begin{eqnarray}
&& \overline{X^0}_R (E^0_L \phi^0 - E^-_L \phi^+), ~~~ 
\overline{X^0}_L (E^0_R \phi^0 - E^-_R \phi^+), \\  
&& (\overline{E^0_L} \phi^+ + \overline{E^-_L} \phi^0) X^-_R, ~~~ 
(\overline{E^0_R} \phi^+ + \overline{E^-_R} \phi^0) X^-_L. 
\end{eqnarray}
The $2 \times 2$ Dirac mass matrix linking $(E^-_L, X^-_L)$ to 
$(E^-_R, X^-_R)$ is of the form
\begin{equation}
{\cal M} = \pmatrix{m_E & m_{21} \cr m_{12} & m_{X^-}},
\end{equation}
and that linking $(E^0_L, X^0_L)$ to $(E^0_R, X^0_R)$ is of the form
\begin{equation}
{\cal M}' = \pmatrix{m_E & m'_{21} \cr m'_{12} & m_{X^0}}.
\end{equation}
In the above, $X^0$ will be assumed to be the lightest, and $X^-$ the 
heaviest. In that case, $X^-$ decays to $E^- h$ ($h$ being the SM Higgs 
boson), $E^-$ decays to $X^0 W^-$ from $E^0-X^0$ mixing, and $E^0$ decays to 
$X^0 h$.   For $y=-2$, the new fermions have $B=1$, with $X^0$ decaying to 
$u_R d_R d_R$ through $\eta$.  For $y=-1$, the new fermions have $B=-1$, 
with $X^0$ decaying to $\bar{u}_R \bar{d}_R \bar{d}_R$ through $\eta^*$. 
Instead of a separate sector, the new fermions are now part of the quarks, 
sharing with them a well-defined global baryon number which is strictly 
conserved.  This connection is made possible by the addition of the 
color scalar diquark $\eta$.

\noindent \underline{\it Scalar Sector}~:~
In addition to the SM Higgs doublet $\Phi$ and the singlet $\zeta$ with 
$B=3$, a scalar singlet $\sigma$ with $B=2$ is added.  For $y=-2$, the term 
$\sigma^* X^0_L X^0_L$ is allowed; for $y=-1$, it is $\sigma X^0_R X^0_R$. 
Hence $\sigma$ decays always to six quarks.  For heavy $\eta$, this 
decay rate will be small enough for $\sigma$ to be long-lived dibaryonic 
dark matter.  The scalar potential consisting of $\Phi$, $\zeta$, and 
$\sigma$ is given by
\begin{eqnarray}
V &=& -\mu_0^2 \Phi^\dagger \Phi -\mu_1^2 \zeta^* \zeta + m_2^2 
\sigma^* \sigma \nonumber \\  
&+& {1 \over 2} \lambda_0 (\Phi^\dagger \Phi)^2 + {1 \over 2} \lambda_1 
(\zeta^* \zeta)^2 + {1 \over 2} \lambda_2 (\sigma^* \sigma)^2 \nonumber \\ 
&+& \lambda_{01} (\Phi^\dagger \Phi)(\zeta^* \zeta) + \lambda_{02} 
(\Phi^\dagger \Phi)(\sigma^* \sigma) + \lambda_{12} (\zeta^* \zeta)
(\sigma^* \sigma).
\end{eqnarray}
Note that because of their $B$ assignments, it is not possible to 
combine $\zeta$ with $\sigma$ in a term with dimension less than five, 
other than $\lambda_{12}$. Let
\begin{equation}
\Phi = \pmatrix{0 \cr (v+h)/\sqrt{2}}, ~~~ \zeta = {1 \over \sqrt{2}} 
(u + H),
\end{equation}
then $v$ and $u$ are determined by the minimum of $V$:
\begin{eqnarray}
0 &=& -\mu_0^2 + {1 \over 2} \lambda_0 v^2 + {1 \over 2} \lambda_{01} u^2, \\ 
0 &=& -\mu_1^2 + {1 \over 2} \lambda_1 u^2 + {1 \over 2} \lambda_{01} v^2, 
\end{eqnarray}
and the $2 \times 2$ mass-squared matrix spanning $h,H$ is
\begin{equation}
{\cal M}^2_{hH} = \pmatrix{\lambda_0 v^2 & \lambda_{01} vu \cr \lambda_{01} vu 
& \lambda_1 u^2}.
\end{equation}
The mass of $\sigma$ is given by
\begin{equation}
m^2_\sigma = m_2^2 + {1 \over 2} \lambda_{02} v^2 + {1 \over 2} \lambda_{12} 
u^2.
\end{equation}
The interactions of $V$ are then
\begin{eqnarray}
V_{int} &=& {1 \over 2} \lambda_0 v h^3 + {1 \over 2} \lambda_1 u H^3 + 
{1 \over 2} \lambda_{01} vhH^2 + {1 \over 2} \lambda_{01} uHh^2 \nonumber \\ 
 &+& \lambda_{02} v h \sigma^* \sigma + \lambda_{12} 
uH\sigma^* \sigma + {1 \over 8} \lambda_0 h^4 + {1 \over 8} \lambda_1 H^4 
\nonumber \\ &+& {1 \over 2} \lambda_2 (\sigma^* \sigma)^2 + {1 \over 4} 
\lambda_{01} h^2 H^2 + {1 \over 2} \lambda_{02} h^2 \sigma^* \sigma + 
{1 \over 2} \lambda_{12} H^2 \sigma^* \sigma.
\end{eqnarray}

The SM gauge bosons $W^\pm$ and $Z$ have their masses in the usual way, 
whereas the $Z_B$ gauge boson becomes massive through $u$, i.e. 
\begin{equation}
m_W = {ev \over 2 \sin \theta_W}, ~~~ m_Z = {ev \over \sin 2 \theta_W}, ~~~ 
m_{Z_B} = 3g_B u. 
\end{equation}
Note that $Z$ and $Z_B$ do not mix.  From collider data, the typical limit on 
a new gauge boson is a few TeV.

\noindent \underline{\it Direct Search of $\sigma$}~:~
The long-lived dark-matter candidate $\sigma$ interacts with quarks through 
$Z_B$ as well as $h$, which must be suppressed to be consistent with 
direct-search experiments by elastic recoil against nuclei in underground 
laboratories.  The effective interaction from $Z_B$ exchange is
\begin{equation}
{\cal L}_{eff} = {2ig_B [\sigma \partial_\mu \sigma^*-\sigma^* \partial_\mu 
\sigma] \over m^2_{Z_B}} \left( {g_B \over 3} \right) \bar{q} \gamma^\mu q.
\end{equation}
The spin-independent elastic scattering cross section of $\sigma$ off a 
xenon nucleus per nucleon is then given by
\begin{equation}
\sigma_0 = {1 \over \pi} \left( {m_\sigma m_{Xe} \over m_\sigma + m_{Xe}} 
\right)^2 \left( {2 \over 9 u^2} \right)^2.
\end{equation}
For $m_\sigma = 1$ TeV, and using $m_{Xe} = 122.3$ GeV with 
$\sigma_0 < 10^{-45}$ cm$^2$~\cite{xenon18}, the lower limit on $u$ is 
92.3 TeV. From Higgs exchange, the analogous cross section is
\begin{equation}
\sigma_0 = {1 \over \pi} \left( {m_\sigma m_{Xe} \over m_\sigma + m_{Xe}} 
\right)^2 \left| {54 f_p + 77 f_n \over 131} \right|^2,
\end{equation}
where~\cite{hint11}
\begin{eqnarray}
{f_p \over m_p} &=& \left[0.075 + {2 \over 27}(1-0.075) \right] 
{\lambda_{02} \over m_h^2 m_\sigma}, \\
{f_n \over m_n} &=& \left[0.078 + {2 \over 27}(1-0.078) \right]
{\lambda_{02} \over m_h^2 m_\sigma}.
\end{eqnarray}
For $m_\sigma = 1$ TeV, $m_h = 125$ GeV, the upper limit on 
$\lambda_{02}$ is then $3 \times 10^{-3}$.  

\noindent \underline{\it Relic Abundance of $\sigma$}~:~
In the early Universe, $\sigma$ interacts with quarks through $Z_B$ and 
remains in thermal equilibrium.  To be a viable dark-matter candidate, 
it must annihilate to lighter particles with the correct interaction 
strength.  Its cross section to quarks through $Z_B$ exchange is proportional 
to $m^2_\sigma/u^4$ which is too small from the direct-search constraint 
$u > 92.3$ TeV.  Two remaining possible choices are $h$ and $H$.  The 
former cross section is proportional to $\lambda_{02}^2$ which is also too 
small, again from the direct-search constraint 
$\lambda_{02} < 3 \times 10^{-3}$. 

As for $H$, the first requirement is that $m_H < m_\sigma$.  Let 
$m_\sigma = 1$ TeV, $m_H = 800$ GeV, and $u = 100$ TeV, then 
$\lambda_1 = 6.4 \times 10^{-5}$ in Eq.~(16).  To suppress $h-H$ mixing, 
this implies $\lambda_{01} << 0.025$.  In this scenario, the cross section 
$\times$ relative velocity of $\sigma \sigma^* \to HH$ is dominated by 
the $t-$channel exchange of $\sigma$, i.e.
\begin{equation}
\sigma_{ann} \times v_{rel} = {\sqrt{1-(m_H/m_\sigma)^2} \over 16 \pi 
m^2_\sigma} \left( {\lambda_{12}^2 u^2 \over 2m^2_\sigma - m_H^2} \right)^2.
\end{equation}
Setting this equal to $3 \times 10^{-26}~{\rm cm}^3/{\rm s}$ for the 
correct relic abundance of dark matter, $\lambda_{12} = 8 \times 10^{-3}$ 
is obtained.  Now $\lambda_{12} \lambda_{01}/\lambda_1 < 3 \times 10^{-3}$ 
from direct search, hence $\lambda_{01} < 2.4 \times 10^{-5}$.  As the 
Universe cools below the temperature around 1 TeV, $\sigma$ freezes out, 
but $H$ decays away quickly to $hh$ through $\lambda_{01} u$.

\noindent \underline{\it Long-Lived Decay of $\sigma$}~:~
The decay to six quarks through two heavy scalars is shown in Fig.~1.
\begin{figure}[htb]
\vspace*{-3.5cm}
\hspace*{-3cm}
\includegraphics[scale=1.0]{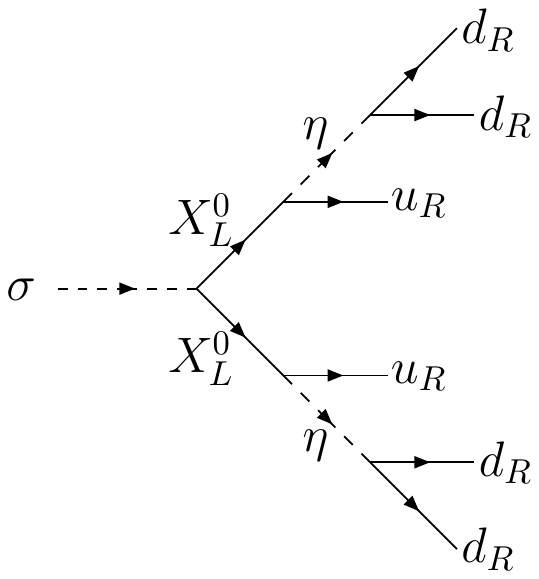}
\vspace*{-21.5cm}
\caption{Decay of $\sigma$ to six quarks}
\end{figure}

\noindent Assuming $X^0$ not to be much heavier than $\sigma$, with $\eta$ 
very heavy compared to them, the undoubtedly overestimate of this decay 
rate is
\begin{equation}
\Gamma \sim \left( {f^2 \over 16 \pi} \right)^5 {m_\sigma^{9} \over 
m^{8}_{\eta}}.
\end{equation}
It should correspond to a lifetime exceeding $10^{25}$~s, not to 
disrupt~\cite{sw17} the observed structure of the cosmic microwave 
background (CMB).  Let $f^2/16 \pi = 10^{-3}$, then for $m_\sigma = 1$ TeV, 
$m_\eta > 4.5 \times 10^7$ GeV.  Generic scalar diquarks are easily produced 
at the Large Hadron Collider (LHC) and observed as energetic quark dijets.  
The current phenomenological lower bound~\cite{psl20} of their mass is 
about 7 TeV.

\noindent \underline{\it Conclusion}~:~
The notion of baryon number $B$ is extended to include a color triplet 
scalar $\eta$ with $B=2/3$ and new fermions with $B=1$ or $B=-1$.  This 
allows the gauging of $B$ which is spontaneously broken by three units, 
with breaking scale above 100 TeV.  A residual global $B$ symmetry remains.  
Assuming now a singlet complex scalar $\sigma$ with $B=2$, it becomes a 
suitable long-lived dark-matter candidate which decays to six quarks.  It 
annihilates to the singlet scalar associated with the breaking of gauged 
$U(1)_B$ for the correct dark-matter relic abundance. It may be observed 
in direct-search experiments through its interactions with the $U(1)_B$ 
gauge boson or the SM Higgs boson.

\noindent \underline{\it Acknowledgement}~:~
This work was supported 
in part by the U.~S.~Department of Energy Grant No. DE-SC0008541.  

\baselineskip 18pt

\bibliographystyle{unsrt}

\end{document}